\documentclass[letter]{aa}  

\usepackage{graphicx}
\usepackage{txfonts}
\usepackage[pdfpagelabels=false]{hyperref}	
\hypersetup{colorlinks=true,linkcolor=blue,citecolor=blue,filecolor=blue,urlcolor=blue,draft=false}
\usepackage{xspace}
\newcommand{\hi}{\ion{H}{I}\xspace}
\newcommand{\hii}{\ion{H}{II}\xspace}

\newcommand{\Mvir}{\ifmmode {M_{\rm vir}} \else $M_{\rm vir}$\xspace\fi}
\newcommand{\Mstar}{\ifmmode {M_\star} \else $M_{\star}$\xspace\fi}
\newcommand{\Msun}{\ifmmode {\rm M}_{\sun} \else ${\rm M}_\sun$\xspace\fi}
\newcommand{\fesc}{\ifmmode {f_{\rm esc}} \else $f_{\rm esc}$\xspace\fi}
\newcommand{\MUV}{\ifmmode {{\rm M}_{\rm UV}} \else ${\rm M}_{\rm UV}$\xspace\fi}
\newcommand{\lya}{\ifmmode {{\rm Ly}\alpha} \else ${\rm Ly}\alpha$\xspace\fi}
\newcommand{\delphi}{\textsc{Delphi}\xspace}
\newcommand{\lzlcs}{\textsc{LzLCS}\xspace}

\begin{document}

   \title{Reionization with star-forming galaxies: insights from the Low-$z$ Lyman Continuum Survey}
   \titlerunning{Modelling reionization with \lzlcs \fesc}

   \author{M. Trebitsch\inst{1}
     \and P. Dayal\inst{1}
     \and J. Chisholm\inst{2}
     \and S. L. Finkelstein\inst{2}  
     \and A. Jaskot\inst{3}
     \and S. Flury\inst{4}
     \and D. Schaerer\inst{5,6}
     \and H. Atek\inst{7}
     \and S. Borthakur\inst{8}
     \and H. Ferguson\inst{9}
     \and F. Fontanot\inst{10,11}
     \and M. Giavalisco\inst{4}
     \and A. Grazian\inst{12}
     \and M. Hayes\inst{13}
     \and F. Leclercq\inst{2}
     \and V. Mauerhofer\inst{1}
     \and G. Östlin\inst{13}
     \and A. Saldana-Lopez\inst{6} 
     \and T. Thuan\inst{14}
     \and B. Wang\inst{15}
     \and G. Worseck\inst{16}
     \and X. Xu\inst{17}
   }

   \institute{Kapteyn Astronomical Institute, University of Groningen, P.O. Box 800, 9700 AV Groningen, The Netherlands\\
     \email{m.trebitsch@rug.nl}
     \and Department of Astronomy, The University of Texas at Austin, Austin, TX, USA
     \and Department of Astronomy, Williams College, Williamstown, MA 01267, United States
     \and Department of Astronomy, University of Massachusetts Amherst, Amherst, MA 01002, United States
     \and Department of Astronomy, University of Geneva, 51 Chemin Pegasi, 1290 Versoix, Switzerland
     \and CNRS, IRAP, 14 Avenue E. Belin, 31400 Toulouse, France
     \and Institut d'Astrophysique de Paris, CNRS UMR7095, Sorbonne Université, 98bis Boulevard Arago, F-75014 Paris, France
     \and School of Earth \& Space Exploration, Arizona State University, Tempe, AZ 85287, United States
     \and Space Telescope Science Institute, 3700 San Martin Drive Baltimore, MD 21218, United States
     \and INAF–Osservatorio Astronomico di Trieste, Via G.B. Tiepolo, 11, I-34143, Trieste, Italy
     \and IFPU–Institute for Fundamental Physics of the Universe, via Beirut 2, I-34151, Trieste, Italy
     \and INAF–Osservatorio Astronomico di Padova, Vicolo dell’Osservatorio 5, I-35122, Padova, Italy
     \and The Oskar Klein Centre, Department of Astronomy, Stockholm University, AlbaNova, SE-10691 Stockholm, Sweden
     \and Astronomy Department, University of Virginia, Charlottesville, VA 22904, United States
     \and Department of Astronomy \& Astrophysics, The Pennsylvania State University, University Park, PA 16802, USA
     \and Institut für Physik und Astronomie, Universität Potsdam, Karl-Liebknecht-Str. 24/25, D-14476 Potsdam, Germany
     \and Department of Physics and Astronomy, Johns Hopkins University, Baltimore, MD 21218, United States
   }

   \date{Received XXX; accepted YYY}
   
   \abstract
   {The fraction of ionizing photons escaping from galaxies, \fesc, is at the same time a crucial parameter in modelling reionization and a very poorly known quantity, especially at high redshift.}
   {Recent observations are starting to constrain the values of \fesc in low-$z$ star-forming galaxies, but the validity of this comparison remains to be verified.}
   {Applying at high-$z$ the empirical relation between \fesc and the UV slope trends derived from the \emph{Low-$z$ Lyman Continuum Survey}, we use the \delphi semi-analytical galaxy formation model to estimate the global ionizing emissivity of high-$z$ galaxies, which we use to compute the resulting reionization history.}
   {We find that both the global ionizing emissivity and reionization history match the observational constraints. Assuming that the low-$z$ correlations hold during the epoch of reionization, we find that galaxies with $-16 \lesssim \MUV \lesssim -13.5$ are the main drivers of reionization. We derive a population-averaged $\langle \fesc \rangle \simeq 8\%, 10\%, 20\%$ at $z=4.5, 6, 8$.}
   {}


   \maketitle
%

\section{Introduction}

Over the past decade, large efforts have been made to understand the epoch of reionization (EoR) \citep[see e.g. the recent reviews of][]{Dayal2018, Robertson2022}, during which the Lyman Continuum (LyC) photons produced by early sources ionized the neutral intergalactic medium (IGM) in around one billion years, with most of the hydrogen being reionized by $z \lesssim 6$ \citep[e.g.][but see also e.g. \citealt{Bosman2022}]{Fan2006}. One of the key questions pertaining to the EoR is the nature of the sources of LyC photons, active galactic nuclei (AGN) or massive stars in star-forming galaxies. While deep observations start to characterize the faint AGN population, observations \citep[e.g.][]{Fontanot2012,Matsuoka2018,Kulkarni2019} and simulations \citep[e.g.][]{Qin2017,Eide2020,Dayal2020,Trebitsch2021,Yung2021} suggest that their role is sub-dominant \citep[but see][]{Giallongo2015,Grazian2018,Grazian2022}, requiring the fainter but more numerous star forming galaxies to be the drivers of reionization.

One key parameter of their contribution is the fraction of LyC photons (\fesc) that can escape after transfer through the interstellar medium (ISM). Unfortunately, this parameter cannot be measured directly during the EoR as the mean free path of LyC photons is too short at $z \gtrsim 4$ \citep{Inoue2014}.
The only way to directly measure \fesc is to use galaxies that form analogues of the sources of reionization, but at lower redshifts where LyC photons can reach us. Using ground-based optical data, several groups have been exploring analogues at $z \gtrsim 3$ \citep[e.g.][although the attenuation by the IGM is not negligible at these redshifts]{Marchi2017, Steidel2018, Vanzella2018, Fletcher2019}.
Alternatively, space-based observations in the ultraviolet (UV) with the Hubble Space Telescope (HST) of $z \sim 0.3$ compact star-forming galaxies sharing similarities with the $z \gtrsim 6$ population \citep[e.g.][]{Schaerer2022} has proven extremely successful in the past few years in detecting LyC emitters \citep[LCEs, e.g.][]{Izotov2016a,Izotov2016b,Izotov2018a,Izotov2018b,Wang2019,Izotov2021,Flury2022} and exploring their physical properties \citep[e.g.][]{Verhamme2017, Gazagnes2020, Ramambason2020, Flury2022b, Xu2022}.
In particular, the recent Low-$z$ Lyman-Continuum Survey \citep[\lzlcs; PI: Jaskot, HST Project ID: 15626,][]{Flury2022} represent a large effort to extend the parameter space of galaxy properties probed by previous LCE samples.
The \lzlcs is a large HST program targeting 66 star-forming $z \sim 0.3$ galaxies over 134 orbits using the Cosmic Origin Spectrograph (COS). Of these galaxies, 35 have a strong $>2\sigma$ LyC detection with \fesc in the range $0.1\% - 20\%$, bringing the number of detected LCEs to 50 out of 89 targeted galaxies when combined with archival data.

We therefore need to assess how well low-$z$ analogues represent the sources of reionization.
Cosmological simulations of galaxies have proved very useful in the past few years: detailed radiation hydrodynamics (RHD) simulations predict a complex, time-dependent, behaviour of \fesc as a function of the host galaxy properties \citep[e.g.][]{Paardekooper2015, Rosdahl2018, Barrow2020}, suggesting in particular that the presence of stellar feedback (both radiative and mechanical) is crucial in allowing LyC photons to reach the IGM \citep[e.g.][]{Wise2014, Kimm2014, Trebitsch2017}. 
However, these simulations are extremely expensive, using tens of millions of computing hours to describe the high-$z$ population alone.
Additionally, cosmological RHD simulations are only starting to reach the resolution needed to model star-forming clouds, a requirement to address this question properly \citep[e.g.][]{Howard2018, Kimm2019, Kimm2022}. It is therefore currently unfeasible to follow down to low redshift the properties of a large sample of galaxies at the appropriate resolution in a full RHD simulation.

In this Letter, we take an alternative approach to test the viability of using low-$z$ samples to probe the physics of the sources of reionization. We model reionization using a semi-analytical galaxy formation model, \delphi \citep{Dayal2014,Dayal2022}, which we combine with a correlation empirically derived from the \lzlcs that relates the slope of the stellar continuum in the UV, $\beta$, and the escape fraction of LyC photons, \fesc. This allows us to extend the low-$z$ LCE properties to high-$z$.


\section{Coupling \delphi to the LzLCS}
\label{sec:methods}

\subsection{Galaxy formation model}
\label{sec:methods:delphi}

We model galaxy formation with the \delphi semi-analytic model \citep{Dayal2014, Dayal2022}. In brief, \delphi uses a binary merger tree approach to jointly track the build-up of dark matter halos and their baryonic components (gas, stellar, metal and dust mass). This model follows the assembly histories of galaxies with halo masses in the range $\Mvir = 10^{8} - 10^{14}\,\Msun$ from $z \sim 40$ down to $z = 4.5$. The available gas mass (from both mergers and IGM accretion) can form stars with an effective star formation efficiency $f_*^{\rm eff} = \min[f_*^{\rm ej}, f_*]$, which is the minimum between the efficiency that produces enough SNII energy to eject the remainder of the gas ($f_*^{\rm ej}$) and an upper maximum threshold ($f_*$) which is one of our free parameters.
In ionized regions, the UV radiation will suppress star formation e.g. by photo-evaporating low-mass haloes or reduce their gas inflows \citep[e.g.][for recent simulations]{Dawoodbhoy2018, Katz2019, Ocvirk2020}.
We model the coupling between the ionizing UV background and galaxy formation as in \citet{Dayal2020} by assuming full photo-evaporation of gas in low-mass haloes with $\Mvir \leq M_{\rm crit}$, preventing them from forming stars. In this work, we explore the case of no UV suppression as well as $M_{\rm crit} = 10^{8.5}, 10^9, 10^{9.5}\,\Msun$, with the intermediate value being our fiducial case. The choice of this value is motivated by the results of the RHD \textsc{CoDa} simulation \citep{Ocvirk2020} and the semi-numerical \textsc{Astraeus} model \citep{Hutter2021}, where haloes below that mass show a suppressed star formation.
We do not include AGN in this model as we explored their role within the \delphi framework in \citet{Dayal2020}.

The dust model is described in \citet{Dayal2022}, so we only summarize its main features here. We include the key processes of production, astration, destruction of dust into metals, ejection, and dust grain growth in the ISM (that leads to a corresponding decrease in the metal mass) to calculate the total dust mass $M_d$ and metal mass $M_Z$ for each galaxy. The model predicts a dust-to-metal ratio of the order of $30\%-40\%$ for our galaxies, consistent with low-metallicity DLA observations \citep{DeCia2013,Wiseman2017}.
We use this dust mass to infer the dust optical depth to UV continuum by assuming that the dust and gas are co-spatially distributed in a disk of radius $r_g = 4.5\lambda r_{\rm vir}$ \citep[e.g.][]{Ferrara2000}, where $r_{\rm vir}$ is the virial radius and $\lambda = 0.04$ \citep[e.g.][]{Davis2009} is the spin parameter of the halo.
Assuming a dust grain size of $a = 0.05\,\mu\mbox{m}$ and a mass density of $s = 2.25\,\mbox{g}.\mbox{cm}^{-3}$ appropriate for a mixture of graphite and carbonaceous grains \citep{Todini2001, Nozawa2003}, this leads to a dust optical depth $\tau_d = 3 M_d/(4\pi r_{\rm gas}^2 a s)$ with a dust extinction efficiency $Q_{\rm ext} = 1$ at $1500\,\AA$. Assuming that stars, gas, and dust are intermixed in the disk locally modelled as a slab, we can write the fraction of UV that is absorbed by the disk as $1 - f_c$, where
\begin{equation}
  \label{eq:fcdust}
  f_c = \frac{1 - e^{-\tau_d}}{\tau_d}.
\end{equation}

Our model contains only two mass- and redshift-independent parameters to match observations: the maximum (instantaneous) star formation efficiency of $f_* = 8\%$ and the fraction $f_w (\approx 7.5\%$) of the SNII energy available to drive an outflow. These parameters have been tuned to simultaneously reproduce the $z \sim 5-12$ stellar mass functions \citep{Gonzalez2011, Duncan2014, Song2016} and UV luminosity function \citep[][LF]{Castellano2010, Mclure2013, Atek2015, Finkelstein2015, Bouwens2016, Calvi2016, Bowler2017,  Livermore2017, Ishigaki2018, Oesch2018, Bouwens2021, Harikane2022}, with the luminosity (UV and ionizing) being inferred from \textsc{Starburst99} templates \citep{Leitherer1999} using the metallicity and the star formation history using a \citet{Kroupa2001} initial mass function between $0.1 - 100\,\Msun$, after attenuation by dust using Eq.~\ref{eq:fcdust}.

\subsection{Escape fraction}
\label{sec:methods:fesc}
We now describe how we infer the escape fraction \fesc of each galaxy as a function of its (attenuated) UV luminosity.
We follow the results of \citet{Chisholm2022} who used the \lzlcs sample consolidated with a selection of other LCEs taken from the literature \citep{Izotov2016a, Izotov2016b, Izotov2018a, Izotov2018b, Wang2019, Izotov2021} for a total of 89 galaxies.

\begin{figure*}
  \centering
  \includegraphics[width=0.48\linewidth]{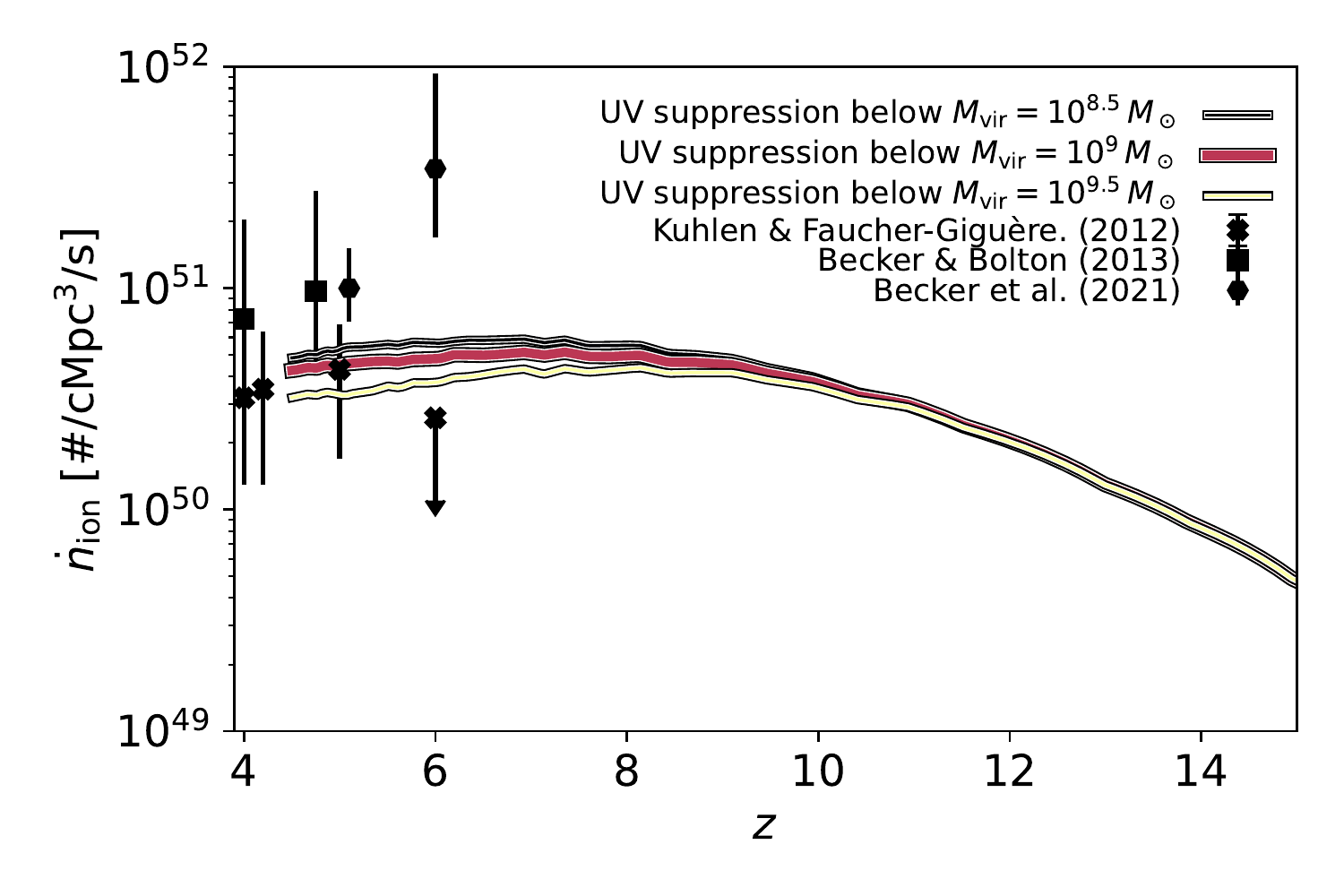}
  \includegraphics[width=0.48\linewidth]{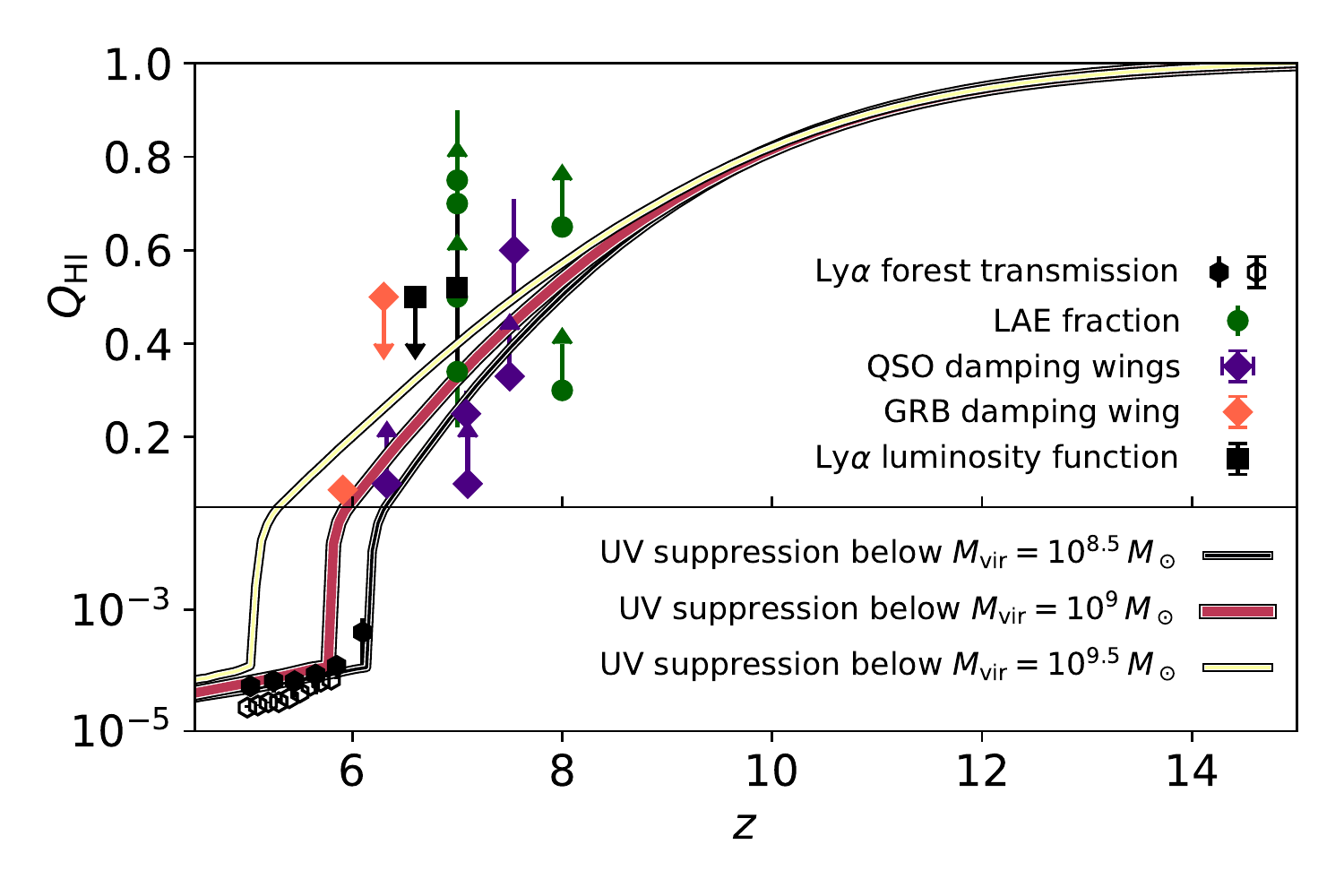}
  \caption{\emph{Left}: Global ionizing photon emission rate for our fiducial model (thick red line) and for the low and high suppression models (yellow and black thin lines, respectively), compared to the observations of \citet[][crosses]{Kuhlen2012} \citet[][squares]{Becker2013}, and \citet[][hexagons]{Becker2021}. \emph{Right}: \hi volume filling fraction for the same models, compared to observations: black hexagons for measurements of the Lyman-$\alpha$ forest transmission from \citet[][full symbols]{Fan2006} and \citet[][open symbols]{Bosman2022}; green circles for constraints on the IGM opacity from the fraction of Lyman-$\alpha$ emitters in Lyman-break galaxy samples \citep[][]{Schenker2014, Ono2012, Pentericci2014, Robertson2013, Tilvi2014}; purple diamonds for measurements from quasar damping wings \citep{Mortlock2011, Schroeder2013, Banados2018, Durovcikova2020}; orange diamonds for gamma-ray bursts constraints \citep[][]{Totani2006, Totani2016}; and the black squares are constraints derived from the evolution of the Lyman-$\alpha$ luminosity function by \citet{Ouchi2010, Ota2008}. Some of these points have been taken from compilation of \citet{Bouwens2015}.
  }
  \label{fig:reionization}%
\end{figure*}

With this extended sample, \citet{Chisholm2022} used the spectral fitting approach of \citet{Saldana-Lopez2022} to reconstruct the stellar populations properties (burst ages and metallicities) of these low-$z$ galaxies, and derived the spectral slope in the range $1300-1800\,\AA$ centred on $1550\,\AA$, $\beta$, from the inferred spectra. Combining $\beta$ with the \fesc measured from the consolidated \lzlcs sample, \citet{Chisholm2022} derived the following relation including non-LCEs as upper limits:
\begin{equation}
  \label{eq:fescbeta}
  \fesc = (1.3 \pm 0.6) \times 10^{-4} \times 10^{(-1.22 \pm 0.1) \beta}.
\end{equation}

We adopt this relation between the LyC \fesc and the UV properties of the galaxies, and proceed to connect these $\beta$ slopes to the UV luminosity of the galaxies predicted by \delphi.
As our goal is to apply this to the high-$z$ population, we choose to adopt a $\beta - \MUV$ relation directly derived from high-$z$ samples rather than the \lzlcs one: indeed, as $\beta$ is strongly related to the dust content of galaxies, the resulting $\fesc - \MUV$ relation will incorporate the properties of high-$z$ dust. While this comes at the cost of full self-consistency, \citet{Chisholm2022} show that the \lzlcs sample follows the same $\beta - \MUV$ trend as at high-$z$.
We use the $\beta - \MUV$ relation of \citet[][see also \citealt{Finkelstein2012, Dunlop2013}]{Bouwens2014}, which can be written as:
\begin{equation}\label{eq:betaMUVdef}
\beta = \beta_{19.5}(z) + \frac{{\rm d}\beta}{{\rm d}\MUV}(\MUV + 19.5),
\end{equation}
where $\beta_{19.5} = \beta(\MUV = -19.5)$.
We fit the evolution of $\beta_{19.5}(z)$ from \citet{Bouwens2014} with a first order polynomial, and fix the slope to $-0.125$, the best-fit constant value:
\begin{equation}\label{eq:betaMUV}
\beta = -1.993 - 0.071(z-6) - 0.125(M_{\rm UV} + 19.5).
\end{equation}
Injecting Eq.~\ref{eq:betaMUV} in Eq.~\ref{eq:fescbeta}, we get a relation between the UV luminosity of a galaxy (from \delphi) and its \fesc. At fixed \MUV, \fesc increases slightly with increasing redshift: from $\simeq 3\%$ ($12\%$) at $z=6$ to $\simeq 6.5\%$ ($19\%$) at $z=10$ for $\MUV = -20$ ($-16$).
We further choose to limit the minimum value of the slope to $\beta = -2.6$, similar to the bluest slopes observed in the \lzlcs sample and consistent with the bluest slopes found at high-$z$ \citep{Finkelstein2012,Bouwens2014}, corresponding to $\fesc \simeq 20\%$.

\subsection{Reionization model}
\label{sec:methods:reionization}

We can now compute the evolution of the volume filling fraction of ionized hydrogen, $Q_{\hii} = 1 - Q_{\hi}$, following the updated ``reionization equation'' approach of \citet[][Eq. 24]{Madau2017}:
\begin{equation}
  \label{eq:madau}
  \frac{\mathrm{d}Q_{\hii}}{\mathrm{d}t} = \frac{\dot{n}_{\rm ion}}{\langle n_{\rm H} \rangle \left(1 + \left\langle \kappa_{\nu_L}^{\rm LLS} \right\rangle / \left\langle \kappa_{\nu_L}^{\rm IGM} \right\rangle \right)} - \frac{Q_\hii}{\bar{t}_{\rm rec}},
\end{equation}
where $\dot{n}_{\rm ion}$ is the emission rate of ionizing photons into the IGM, $\langle n_{\rm H}\rangle$ is the mean density of hydrogen, and $\bar{t}_{\rm rec} = 1/\left[ (1+\chi) \langle n_{\rm H}\rangle \alpha_0 C_R \right]$ is the effective recombination timescale in the IGM. $\bar{t}_{\rm rec}$ depends on $\alpha_0$ the case-A recombination coefficient (taken at $T_0 = 10^{4.3}\,\mbox{K}$) and on the clumping factor $C_R = 2.9 \left[(1+z)/6\right]^{-1.1}$ from \citet{Shull2012}, and $\chi = 0.083$.
Finally, $\left\langle \kappa_{\nu_L}^{\rm LLS} \right\rangle$ and $\left\langle \kappa_{\nu_L}^{\rm IGM} \right\rangle$ are the volume-averaged absorption probabilities per unit length due to Lyman-limit systems and the uniform IGM, respectively \citep[][Eq.~11 and Eq.~13]{Madau2017}.

In this work, we use our \delphi runs described in Sect.~\ref{sec:methods:delphi} to estimate the intrinsic ionizing photon production rate of each galaxy ($\dot{N}^{\rm int}$, in photons/s) and its escape fraction \fesc based on its \MUV. The total ionizing photon emission rate $\dot{n}_{\rm ion}$ is then given by the sum of the contribution coming from all haloes, weighted by the halo mass function $\phi(M)$.
We account for the effect of UV radiation on galaxies in ionized regions by using our ``photo-suppressed'' models for a fraction $Q_{\hii}$ of the galaxies, and the case without UV suppression for the remaining fraction $Q_{\hi}$, as in, e.g. \citet{Choudhury2019}. Formally:
\begin{equation}
  \label{eq:ndotion}
  \dot{n}_{\rm ion} = \int_{\Mvir} \fesc \left[ \dot{N}^{\rm int}_{\rm UVsup} Q_\hii + \dot{N}^{\rm int}_{\rm noUVsup} Q_\hi  \right] \phi(M)\mathrm{d}M.
\end{equation}
This expression depends non-trivially on $Q_\hi$, so that solving Eq.~\ref{eq:madau} effectively takes into account the radiative feedback on the galaxy population. In practice, we solve Eq.~\ref{eq:madau} using a backward differentiation formula method with \textsc{Scipy}'s \texttt{OdeSolver}\footnote{\label{fn:scipy}\url{https://scipy.org/}}.

\begin{figure*}
  \centering
  \includegraphics[width=.48\linewidth]{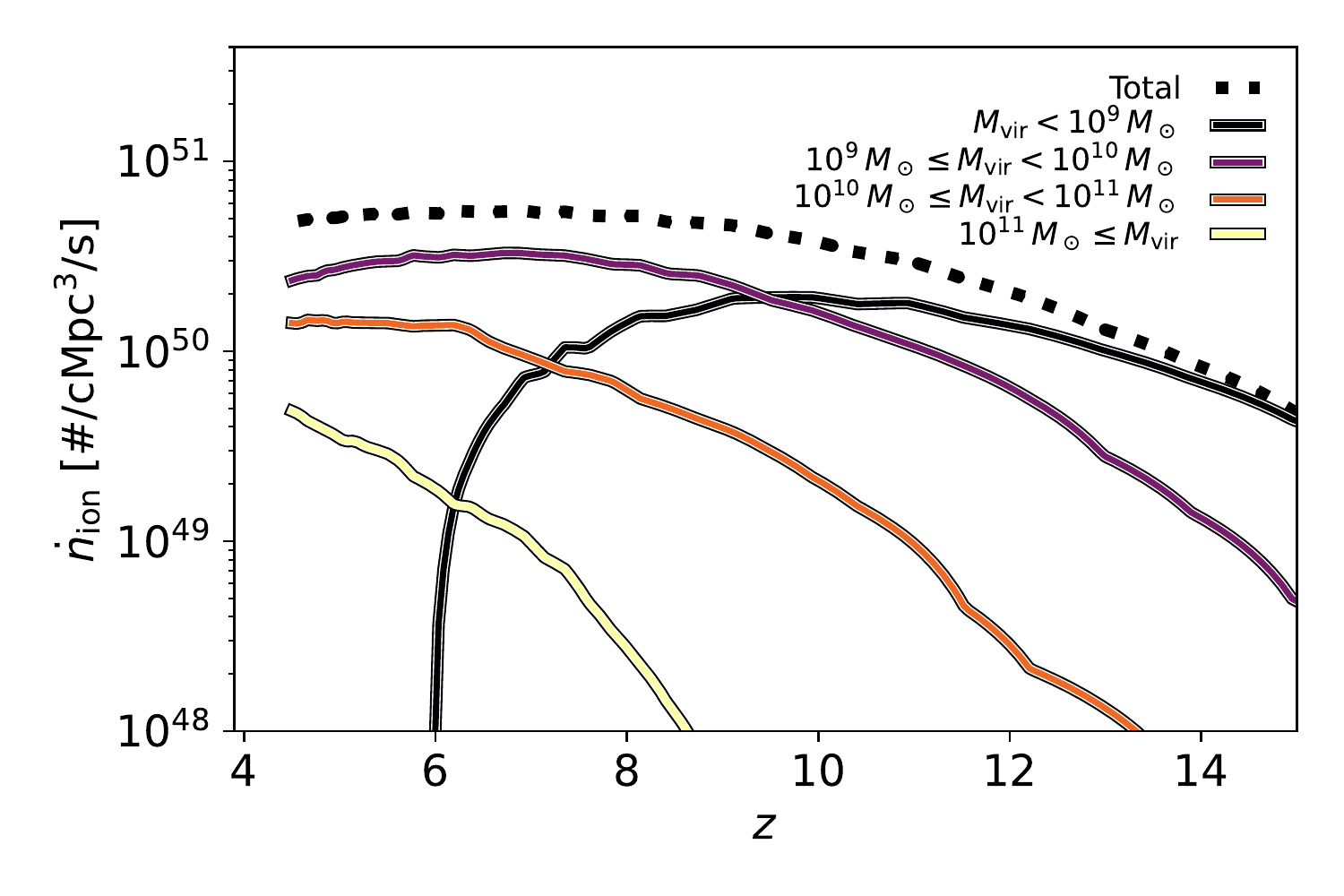}
  \includegraphics[width=.48\linewidth]{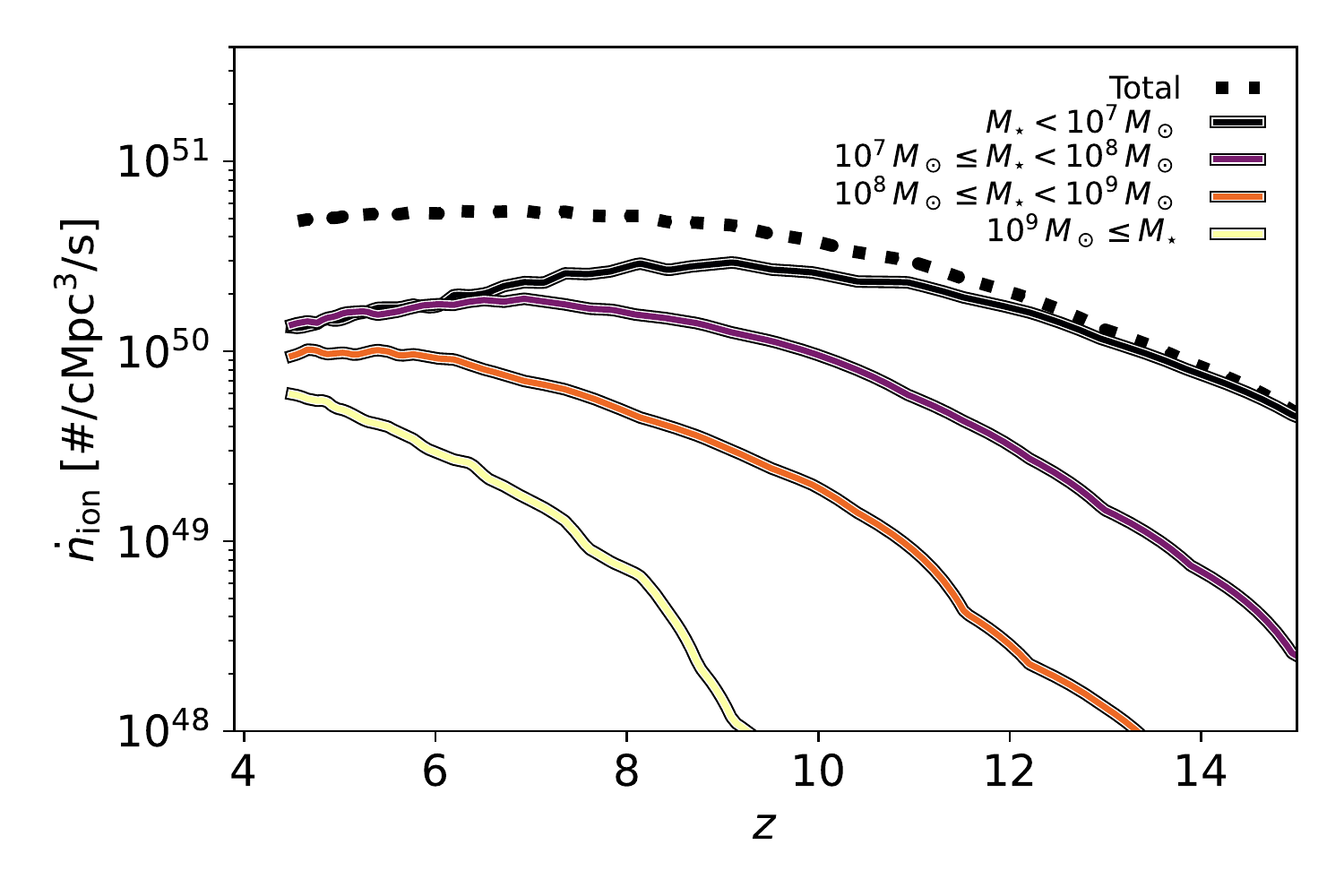}\\[-3ex]
  \includegraphics[width=.48\linewidth]{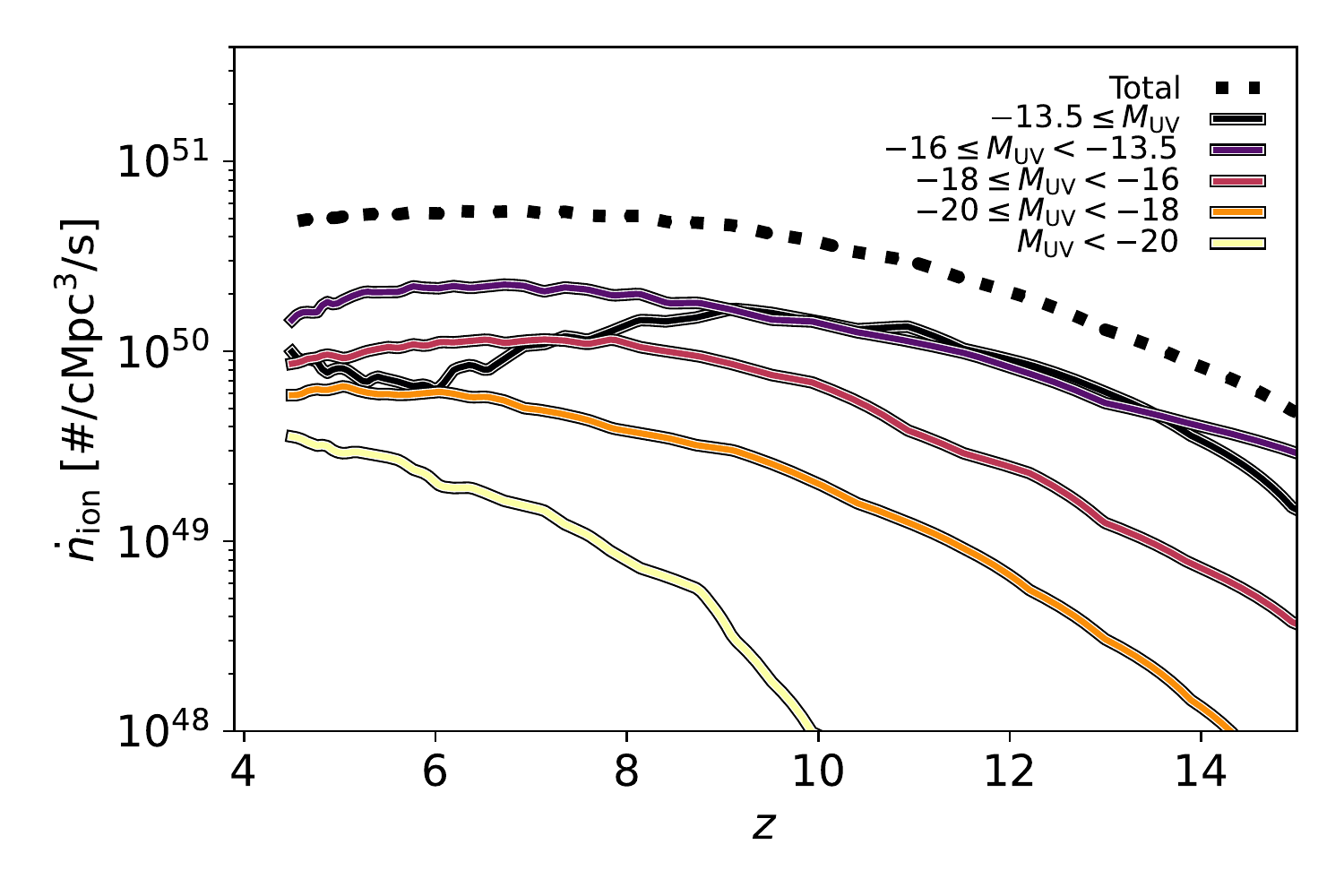}
  \includegraphics[width=.48\linewidth]{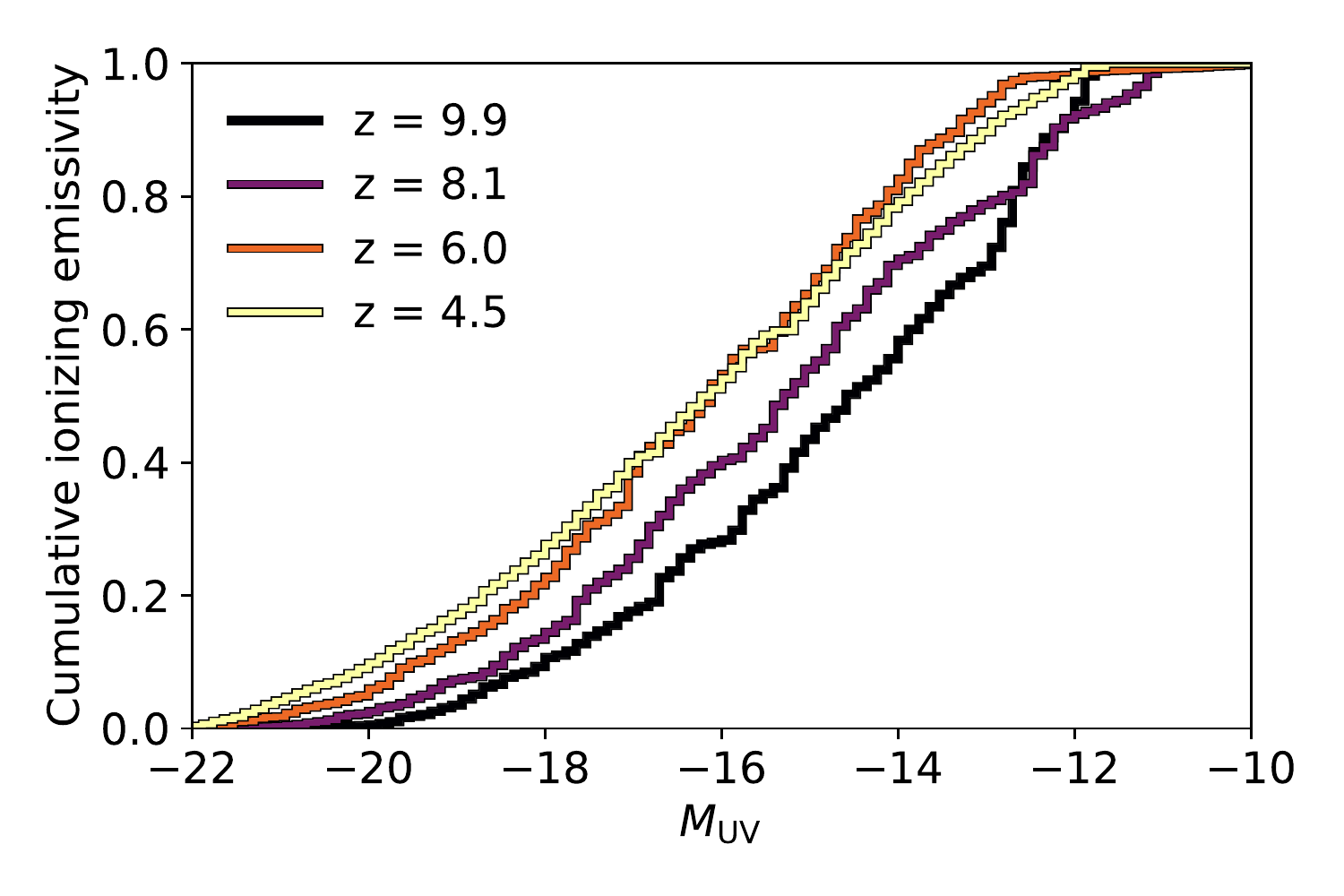}
  \caption{Evolution of the ionizing emission rate for our fiducial model for the whole galaxy population (dotted line) and for haloes of different total mass (\emph{top left}), stellar mass (\emph{top right}), and observed UV magnitude (\emph{bottom left}). The \emph{bottom right} panel shows the cumulative emissivity as a function of observed \MUV. Reionization is driven by relatively low-mass and faint galaxies in haloes with masses $10^{9}\, \Msun \leq \Mvir < 10^{10}\,\Msun$.}
  \label{fig:ndotesc}%
\end{figure*}

\section{Results}
\label{sec:results}

\subsection{Global reionization}
\label{sec:results:global}

We present in Fig.\ref{fig:reionization} the main results from our modelling: the left panel shows the evolution of $\dot{n}_{\rm ion}$ as a function of redshift for our fiducial model (in red), where star formation is suppressed in haloes with $\Mvir \leq 10^9\,\Msun$ in ionized regions, and the models with UV suppression below $10^{8.5}\,\Msun$ and $10^{9.5}\,\Msun$ are shown in black and yellow, respectively. As expected, when the critical mass for UV suppression increases, the total emissivity decreases: this is directly due to the fact that fewer galaxies contribute to the total emissivity, with our most suppressed model producing approximately half as many photons as the least suppressed one at $z \lesssim 8$. The photo-suppression only plays a role in suppressing galaxies when the UV background is well established (i.e. $Q_\hii \simeq 1$), only causing minor differences in $\dot{n}_{\rm ion}$.
All three models predict an emissivity consistent with the observations from \citet[][black squares]{Becker2013}, but on the lower end of the allowed range: this is most likely because we do not include AGN in this model, while they start to make a significant contribution at $z \lesssim 4.5$ \citep{Finkelstein2019, Dayal2020, Trebitsch2021}.

The resulting reionization history is shown in the right panel of Fig.~\ref{fig:reionization}, comparing our model to a selection of observational constraints (see caption).
Our fiducial model reproduces best all observational constraints: the model with the strongest UV suppression lacks the ionizing photons to complete reionization in time, while the model with a low critical mass reionizes the Universe just too early. The post-overlap behaviour of our model is mostly dictated by the inclusion of the $\left\langle \kappa_{\nu_L}^{\rm LLS} \right\rangle$ term, designed to reproduce the $z < 5.5$ forest, so the agreement with the \citet{Fan2006} points is not surprising.
Overall, we find that if the LyC trends from the \lzlcs can be extended to high-$z$, our model suggests (without specific tuning) that galaxies alone can reionize the Universe by $z \lesssim 6$.

\subsection{Which galaxies are the main drivers of reionization?}
\label{sec:results:sources}

Having established that our model reproduces reasonably well the global reionization history, we can use it to probe the physics of the sources of reionization. We show in Fig.~\ref{fig:ndotesc} a breakdown of the contributions to $\dot{n}_{\rm ion}$ from different galaxy population for our fiducial model, sorted by halo mass (top left), stellar mass (top right), and observed UV luminosity (bottom left). In all three panels, the dashed line indicate the total $\dot{n}_{\rm ion}$.
The ionizing budget is dominated at early times by the lowest mass haloes, with $\Mvir < 10^9\,\Msun$, but their contribution drops as they are more and more photo-suppressed, becoming negligible at $z \lesssim 6$ when reionization is complete. During the majority of the reionization era ($6 \lesssim z \lesssim 9$), the haloes with masses $10^{9}\, \Msun \leq \Mvir < 10^{10}\,\Msun$ represent the main contributors to $\dot{n}_{\rm ion}$ ($\sim 60\%$ at $z\simeq 8$).
Grouping galaxies by stellar mass yields a similar behaviour: at $z \gtrsim 10$, the galaxies with $\Mstar < 10^7\,\Msun$ dominate as they live in haloes that are not yet photo-suppressed. Their contribution then decreases, but remains dominant: this is mostly because as more massive haloes get partly photo-suppressed, they will host galaxies in that stellar mass bin while contributing to $\dot{n}_{\rm ion}$ \citep[see e.g.][]{Hutter2021}. By the end of reionization, galaxies with masses $10^{7}\, \Msun \leq \Mstar < 10^{8}\,\Msun$ contribute as much as this lowest mass bin, and galaxies with $10^{8}\, \Msun \leq \Mstar < 10^{9}\,\Msun$ contribute almost as much. This suggests an intermediate pathway between the ``reionization by the faintest galaxies'' \citep[e.g.][]{Finkelstein2019} and the ``reionization by rare galaxies'' \citep[e.g.][]{Naidu2020} scenarios.
Finally, we find that at $6 \lesssim z \lesssim 9$ galaxies with $-16 \lesssim \MUV \lesssim -13.5$ drive reionization. By comparison, the galaxies fainter than $\MUV = -13.5$ only play an important role at the beginning of reionization, with their contribution remaining similar to that of galaxies with $-18 \lesssim \MUV \lesssim -16$ and $-20 \lesssim \MUV \lesssim -18$ at later time as a result of UV suppression, and the brightest galaxies remain subdominant throughout the reionization era. We show in the bottom right panel of Fig.~\ref{fig:ndotesc} the cumulative contribution to $\dot{n}_{\rm ion}$ as a function of observed \MUV: the contribution of faint galaxies decreases as reionization progresses, with the contribution from galaxies brighter than $\MUV = -15$ evolving from $\simeq 40\%$ at $z \simeq 10$ to over $60\%$ at $z \lesssim 6$.

This paints a picture that contrasts with the empirical results of e.g. \citet{Naidu2020}, whose model suggests that relatively bright galaxies contribute significantly, while here they only contribute at late time. This discrepancy mainly comes from different assumptions for the behaviour of \fesc with galaxy properties: \citet{Naidu2020} implement a model where more massive galaxies have a higher \fesc, driven by the idea that stronger feedback carves holes more efficiently in the ISM, as advocated e.g. by \citet{Sharma2016}; they further assume a relatively shallow UV LF, leading to a lower contribution of low-luminosity sources. By contrast, our observationally motivated \fesc model yields a higher \fesc for faint, low-mass galaxies, in better agreement with recent RHD simulations \citep[e.g.][]{Katz2019, Lewis2020, Rosdahl2022}. With our model, the bulk of the reionizing population from \citet{Naidu2020} ($\MUV < -18$) has $\beta > -2.2$, and so $\fesc < 8\%$.
Nevertheless, while our model suggest that faint galaxies drive reionization, it is less extreme that the \citet{Finkelstein2019} scenario, where \emph{the faintest} galaxies are responsible for reionization.

\begin{figure}
  \centering
  \includegraphics[width=\linewidth]{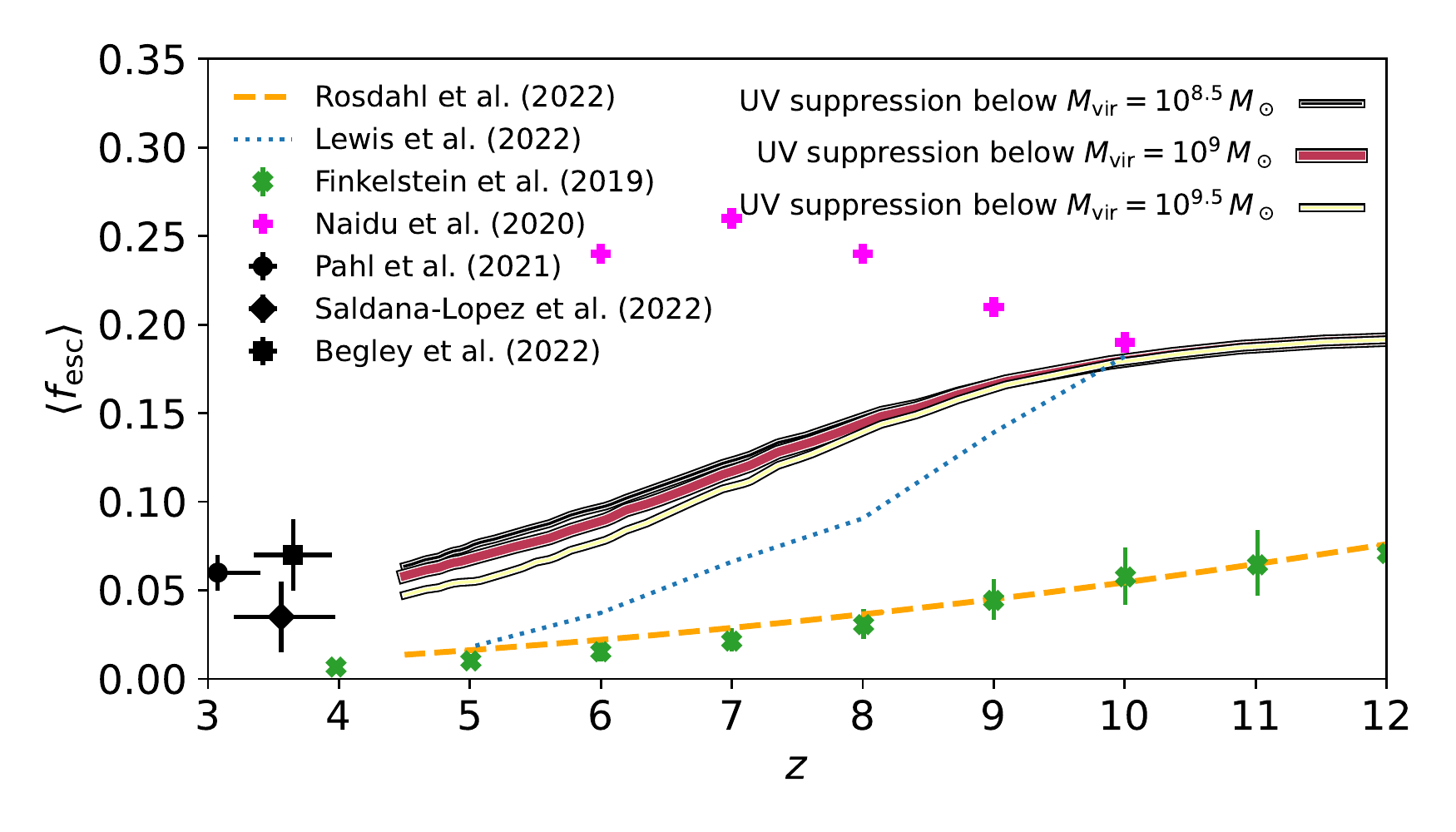}
  \caption{Evolution of the luminosity-weighted, population-averaged \fesc for our three models: we find a slow decrease from $\langle\fesc\rangle \simeq 20\%$ at $z \gtrsim 10$ to $\langle\fesc\rangle \simeq 8\%$ at $z \simeq 4$. The extrapolation of our model to lower redshift aligns well with observational constraints \citep[e.g.][black points]{Pahl2021,Saldana-Lopez2022b,Begley2022}, and is in qualitative agreement with the trend found by models \citep[][green and pink crosses]{Finkelstein2019,Naidu2020} and simulations \citep[][orange dashed line]{Rosdahl2022}, close to the simulation of \citet[][blue dotted line]{Lewis2022}.}
  \label{fig:fesc}
\end{figure}

A key prediction of the \citet{Naidu2020} and \citet{Finkelstein2019} empirical models is the evolution of the global \fesc with redshift. Driven by the evolution of the faint-end slope of the UV luminosity function, \citet{Finkelstein2019} find that \fesc decreases with increasing cosmic time, while \citet{Naidu2020} find the opposite.
We show in Fig.~\ref{fig:fesc} our luminosity-weighted, population-averaged \fesc with the same colour-coding as Fig.~\ref{fig:reionization}. At $z \gtrsim 9-10$, the global $\langle\fesc\rangle$ saturates around $20\%$, our maximum value, before decreasing at lower $z$: this is because the lowest mass haloes dominate the very early emissivity, while they get photo-suppressed as reionization proceeds. Again, we qualitatively agree with \citet{Finkelstein2019}, albeit with quantitative differences: at $z\simeq 4.5, 6, 8$, we find $\langle\fesc\rangle \simeq 8\%, 10\%, 20\%$, compared to $\lesssim 1\%, 2\%, 5\%$ in their model. This slow decrease of $\langle\fesc\rangle$ is in good agreement with $z\simeq 3$ observations and upper limits \citep[e.g.][]{Guaita2016, Steidel2018, Mestric2021, Pahl2021, Begley2022, Saxena2022,Saldana-Lopez2022b} which have $\fesc \lesssim 10\%$ \citep[][but see also \citealt{Rivera-Thorsen2022}]{Begley2022}, suggesting at face value that \lzlcs galaxies exhibit LyC properties broadly similar to the high-$z$ galaxy population. While RHD simulations usually stop before $z \simeq 3$, the trend of $\langle\fesc\rangle$ decreasing with lower redshift is a common feature: \citet{Rosdahl2018,Rosdahl2022} find that $\langle\fesc\rangle$ goes from $\simeq 12-15\%$ at $z\simeq 15$ to $\simeq 3-8\%$ at $z = 6$, in qualitative agreement with our model. Similarly, \citet{Trebitsch2021} find a decreasing $\langle\fesc\rangle$ in the \textsc{Obelisk} simulation down to values close to $\simeq 2-3\%$ at $z\simeq 4$, while \citet{Lewis2022} and the \emph{Alt6} model of \citet{Dayal2020} find an evolution of \fesc with $z$ very similar to ours.

\section{Summary}

In this Letter we have combined the semi-analytical galaxy formation model \delphi with observations of low-$z$ LyC emitters from the (consolidated) \lzlcs. By relating the escape of LyC photons with the UV properties of the host galaxy, we have built a simple but consistent model for the reionization of the Universe. Our main results are as follow:
  \begin{itemize}
  \item A direct application of the $\fesc - \beta$ relation found in the \lzlcs to galaxies in the EoR yield a reasonable population of sources of ionizing photons. Using these sources to solve the reionization equation matches current observational constraints on the reionization history.
  \item The faintest galaxies in the lowest mass haloes dominate the early stages of reionization, but become sub-dominant at $z \lesssim 9$ when they become photo-suppressed by the UV background that becomes more prevalent.
  \item At the end of reionization, galaxies with $-16 \lesssim \MUV \lesssim -13.5$, which will be observable with the \emph{JWST}, are dominating the LyC budget, while the brightest galaxies with $\MUV < -20$ are sub-dominant at all times.
  \item The average $\langle\fesc\rangle$ decreases over time, going from $\langle \fesc \rangle \simeq 20\%$ at $z\gtrsim 8$ to $\langle \fesc \rangle \simeq 8\%$ at $z\simeq 4$, in qualitative agreement with the predictions of high-resolution RHD simulations and observations at $z \gtrsim 3$.
  \end{itemize}
  These results demonstrate the power of using observations of low-$z$ galaxies with a strategy such as that of the \lzlcs to get insights on the population of galaxies deep in the reionization epoch, and pave the way for more detailed models and comparisons with high-$z$ samples.

\begin{acknowledgements}
  MT thanks J. Lewis for sharing results from the DUSTiER simulation.
  MT, PD, and VM acknowledge support from the NWO grant 0.16.VIDI.189.162 (``ODIN''). PD acknowledges support from University of Groningen's CO-FUND Rosalind Franklin Program.
  ASL acknowledge support from Swiss National Science Foundation.
  This work has made extensive use of the NASA's Astrophysics Data System, as well as the \textsc{Matplotlib} \citep{Hunter2007}, \textsc{Numpy/Scipy} \citep{Harris2020} and \textsc{IPython} \citep{Perez2007} packages.
  This research is based on observations made with the NASA/ESA Hubble Space Telescope obtained from the Space Telescope Science Institute, which is operated by the Association of Universities for Research in Astronomy, Inc., under NASA contract NAS 5–26555. These observations are associated with program HST-GO-15626.
\end{acknowledgements}

\bibliographystyle{aa} 
\bibliography{bibliography.bib} 

\end{document}